# Observation of PT-symmetric quantum interference

**Authors:** F.Klauck†, L. Teuber†, M. Ornigotti, M.Heinrich, S. Scheel, A. Szameit*.

**Affiliations:**

[1]Institut für Physik, Universität Rostock, Germany.

*Correspondence to: alexander.szameit@uni-rostock.de

†equal contributions

**Abstract:**

Parity-Time (PT) symmetric quantum mechanics is a complex extension of conventional Hermitian quantum mechanics in which physical observables possess a real eigenvalue spectrum. However, an experimental demonstration of the true quantum nature of PT symmetry has been elusive thus far, as only single-particle physics has been exploited to date. In our work, we demonstrate two-particle quantum interference in a PT-symmetric system. We employ integrated photonic waveguides to reveal that PT-symmetric bunching of indistinguishable photons shows strongly counterintuitive features. We substantiate our experimental results by modelling the system by a quantum master equation, which we analytically solve using Lie algebra methods. Our work paves the way for nonlocal PT-symmetric quantum mechanics as a novel building block for future quantum devices.

**One Sentence Summary:** Counterintuitive photon bunching characteristics in PT-symmetric quantum mechanics.

As a complex extension of conventional Hermitian quantum mechanics, Parity-Time (PT)-symmetric quantum mechanics (*1*) garnered substantial interest in recent years (*2*). As PT-symmetric systems may possess an entirely real eigenvalue spectrum despite being non-Hermitian, they exhibit hallmark features such as non-orthogonal eigenmodes (*3*), exceptional points (*4*), and diffusive coherent transport (*5*). Moreover, PT-symmetry has profound implications in systems including nonlinearity (*6*), lasing (*7*), and non-trivial topologies (*8, 9*). Although PT-symmetry is mostly being explored using photonic platforms, it has enriched other research fields such as atomic diffusion (*10*), superconducting wires (*11*), and electronic circuits (*12*). Despite its general formulation, all experimental demonstrations to date focused on single-particle phenomena that is, on first-quantized quantum mechanics using wavefunctions and their analogies, either in the classical regime (*13*) or the single-photon realm (*14*). However, only in second quantization where the wavefunction itself is quantized, does the true nature of quantum physics make its appearance. Yet, an experimental demonstration of PT-symmetry in second quantization, that is, including nonlocality and non-realism, has so far been elusive.

Here, we demonstrate experimentally two-photon interference in an integrated, PT-symmetric optical structure. Our system consists of two evanescently coupled optical waveguides that implement the integrated analogue of a bulk beam splitter. The seminal work by Hong, Ou and Mandel (*15*) showed that indistinguishable photons exhibit quantum interference resulting in (maximally) entangled states. In contrast to bulk beam splitters, however, waveguides allow for additional degrees of freedom in exploring the underlying physics. For example, individual waveguides can be coupled to their own respective reservoirs, that is, they may experience different loss, something that cannot be achieved in bulk (*16*). On this basis, non-Hermitian systems exhibiting PT-symmetry can be realized. In particular, this allows for the asymmetric coupling of a nonlocal quantum state to an external reservoir (Fig. 1A). By modulating one of the waveguides perpendicular to the propagation direction of the light, effective radiation losses result in coupling to a reservoir and, hence, to Markovian loss (Fig. 1B). This is reflected in a nontrivial imaginary part of the waveguides' refractive indices (*17*) which is known to be a basis for implementing PT-symmetry (*18, 19*). As we show below in theory and experiment, this asymmetry has a profound impact on the characteristics of the photonic quantum interference that is strongly counterintuitive.

We model our system by a quantum master equation in Lindblad form which reads as

$$\partial_z \rho(z) = -\frac{i}{\hbar}[H,\rho] + \gamma\left(2 a_L \rho a_L^\dagger + \{a_L^\dagger a_L, \rho\}_+\right) = \mathcal{L}\rho. \qquad (1)$$

Here, $\rho(z)$ is the photonic density operator as a function of propagation distance $z$. The amplitude operators $a_L$ and $a_R$ are associated with the quantized photonic excitations in waveguides $L$ and $R$. The unitary evolution is governed by the Hamiltonian $H = \kappa(a_L^\dagger a_R + a_L a_R^\dagger)$ with the coupling rate $\kappa$ between waveguides $L$ and $R$. The coupling of waveguide $L$ to the reservoir is modelled by the loss rate $\gamma$. In order to solve the Lindblad master equation, we choose an approach based on a Lie algebra treatment that provides us with an eigendecomposition of the density matrix. To this end, we transform Eq. (1) from Hilbert space to Liouville space (20), which is effectively a reformulation of the Liouvillian $\mathcal{L}$ in terms of left-right actions of the photonic amplitude operators. In this representation, the formal solution of the master equation can be obtained by diagonalizing the Liouvillian. This procedure provides us with the eigenmodes of $\mathcal{L}$, which in turns allows us to write the solution of the master equation as

$$\rho(z) = e^{\mathcal{L}z}\rho_0 \qquad (2)$$

for a given input density operator $\rho_0$. Its eigenmodes can be seen as the simultaneous creation and annihilation of photons in the waveguides $L$ and $R$. For details of the calculation, we refer the interested reader to the Supplementary Material.

With this analytical solution at hand, we study the evolution of the diagonal elements of the density matrix (2) through the lossy (PT-symmetric) waveguide coupler. For reference, in Fig. 2A we show the well-known Hermitian evolution of a single photon in a system of identical lossless waveguides. In this case, at half the coupling length $L_c/2 = \pi/(4\kappa)$, the probabilities for finding that photon in each waveguide are equal. The well-known Hong-Ou-Mandel (HOM) quantum interference of two indistinguishable photons would be observed at this propagation distance. A common misconception regards the exact location at which this HOM dip occurs. In a Hermitian scenario, the propagation distance at which the probability to find a photon in one waveguide amounts to 50 % coincides with the position of the HOM dip. In non-Hermitian situations, however, this is no longer the case. Indeed, as we show in Figs. 2B and 2C, the location of the HOM dip is not directly related to the position of equal photon number probabilities. To demonstrate this surprising effect, in Fig. 2B we depict the diagonal elements of the density matrix of a single photon in a lossy PT-symmetric waveguide coupler. Depending on which waveguide is excited, the location of equal photon number probabilities are different, but never at half the coupling length $L_c/2$. In Fig. 2C, we show the coincidence rate $\Gamma(z)$ for a state of two indistinguishable photons $|1,1\rangle$ where one photon is launched into waveguide $L$ and the other into waveguide $R$. The coincidence rate can be computed from the density matrix and reads

$$\Gamma(z) = e^{-2\gamma z}\left(\frac{\gamma^2 - 4\kappa^2 \cos(\omega z)}{\omega^2}\right)^2, \tag{3}$$

where $\omega = \sqrt{4\kappa^2 - \gamma^2}$ is the oscillation frequency between the waveguides. In the Hermitian case ($\gamma = 0$), the coincidence rate is an oscillating function with frequency $2\kappa$. Therefore, the first root is found at $z_H = \frac{\pi}{4\kappa}$ which is exactly the position of the HOM dip. In the non-Hermitian PT-symmetric case ($\gamma > 0$), the oscillation frequency $\omega$ is smaller. However, the influence of the loss in Eq. (3) results in a root of the coincidence rate $\Gamma(z)$ at $z_0 = \frac{2}{\omega}\arcsin\left(\frac{\omega}{\sqrt{8}\kappa}\right) < z_H$. In other words, the position of the HOM dip, i.e. the location at which both photons are found in the same output and hence form an entangled state (Fig. 2D), shifts to shorter propagation distances, despite the reduced oscillation frequency. If the loss is too large ($\gamma > 2\kappa$), the oscillation frequency becomes imaginary as the PT-symmetry is broken. The shaded area in Fig. 2C indicates the region in which a HOM dip can occur. Only in the lossless case, the HOM dip occurs at $z_H$. In the non-Hermitian case, the HOM dip occurs at $z < z_H$, approaching $z = 1/(\sqrt{2}\kappa)$ for $\gamma \to 2\kappa$. For details of the calculation, see the Supplementary Material.

This is a seemingly counterintuitive behaviour. On the one hand, the loss $\gamma$ reduces the oscillation frequency of the photons between the waveguides. That is, the position at which the photons are equally distributed, depends on the excited waveguide and is shifted with respect to the lossless case. Depending on which waveguide is illuminated, the position of equal photon distribution is shifted towards slightly smaller propagation distances (lossy waveguide $L$ excited) or to much larger distances (lossless waveguide $R$ excited), as shown in Fig. 2B. However, despite the average position of equal photon distribution being at $z > z_H$, the position of the HOM dip is at $z < z_H$. The reason is the unexpected influence of the loss factor $\gamma$ that not only determines the oscillation frequency of the coincidence rate, but also the position of its root.

For our experiments, we fabricate the samples using the femtosecond direct laser writing technique in fused silica glass (*21*). Details regarding the laser writing can be found in the Methods section. Each sample consists of two waveguides with a geometry shown in Fig. 1B. In the experiments, we varied two parameters -- the loss as well as the propagation distance. The loss was implemented by a slight sinusoidal modulation of the left waveguide, resulting in additional tunable bending losses (*22*). For each non-Hermitian sample, a reference Hermitian counterpart was perused.

In a first set of experiments, we characterize the samples using classical laser light at $\lambda = 815$ nm. To this end, we launch the light into one of the waveguides and measure the intensity ratio at the output facet. In this way, we can extract the required information of the dynamics without being affected by the mean loss (which is necessary for a passive PT-symmetric system). The set of samples with varying propagation length but identical loss provide us with full information of the dynamics of the propagation. The results are summarized in Fig. 3. In the Hermitian case ($\gamma = 0$, black data), the intensity ratio is independent of the excited waveguide. The coupling length was determined to be $L_c = 6.0$ cm with the coupling constant $\kappa = 0.26$ cm$^{-1}$. The HOM dip in this case occurs at $z_H = L_c/2 = 3.0$ cm. In the non-Hermitian PT-symmetric case ($\gamma > 0$), however, the intensity ratio depends strongly on the chosen input waveguide. If the lossless waveguide $R$ is excited, the intensity ratio decreases much more slowly compared to the Hermitian situation (blue data), whereas, if the lossy waveguide $L$ is excited, the intensity ratio drops much faster (red data). As a consequence, the point of equal intensity distribution is shifted with respect to the Hermitian case.

In our central experiment, we investigate photon coincidences by launching pairs of indistinguishable photons at $\lambda = 815$ nm into the samples. Information regarding the photon source and the launching procedure can be found in the Methods section. The degree of indistinguishability is controlled by launching the photons with varying time delays into the waveguides, in the spirit of the original HOM experiment (*15*). At zero delay, the quantum interference of the photons is at its most destructive. Performing such a HOM measurement for different propagation lengths in the Hermitian case results in data shown in Fig. 4A. The minima are interpolated using a Gaussian fit. The exact location of the HOM dip with a visibility of (87±2) % is at the minimum coincidence rate for all propagation lengths and time delays. The data for the lossy PT-symmetric situation with $\gamma = (0.13 \pm 0.04)$ cm$^{-1}$ is shown in Fig. 4B where the HOM dip exhibits a visibility of (90 ±4) %. Combining the full set of measurements for all propagation lengths leads to Fig. 4C. While the HOM dip for the Hermitian case (black data) occurs at a propagation distance $z_H = 3.0$ cm, its location is shifted in the non-Hermitian situation (red data) to $z = 2.8$ cm $< z_H$ in full agreement with our theory (solid lines). The observed shift of 0.2 cm clearly exceeds the prediction uncertainty of our fitted theory for the minimum of the HOM dip of 0.05 cm.

In our work, we demonstrate PT-symmetric quantum optics beyond first quantization. We theoretically and experimentally show how loss in passive PT-symmetric optical systems affects quantum optical phenomena such as the Hong-Ou-Mandel quantum interference. We predict and observe a counterintuitive shift of the position of the HOM dip in integrated lossy waveguide structures to shorter propagation distances. Our results present an interesting conundrum: the observation of the coincidence rate necessitates that both photons are not lost during transmission through the waveguides. Nonetheless, the very fact that the photons could have been lost, results in a significant modification of the quantum dynamics. In other words, the mere possibility of losing a photon already changes the propagation dynamics. This observation poses a number of interesting questions. First, can a similar approach as described in our work using losses be adapted to nonlinear settings, that is, in which the photon dynamics evolves according to a nonlinear interaction despite not having had contact with a nonlinear medium? Second, as PT-symmetric Hamiltonians form a class of operations that only partially overlap with Hermitian operators, one might envision the design of novel quantum gates based on PT-symmetry that would not be accessible in a Hermitian framework. Third, how does the concept of exceptional points extend to second quantization? These and other questions can now be answered using our experimental platform.

## Funding

The authors acknowledge funding from the Deutsche Forschungsgemeinschaft (grants SCHE 612/6-1, SZ 276/12-1, BL 574/13-1, SZ 276/15-1, SZ 276/20-1) and the Alfried Krupp von Bohlen and Halbach foundation. This project has received funding from the European Union's Horizon 2020 research and innovation programme under grant agreement No 800942.

## Authors Contributions
F. K. and A.S. developed the idea. L.T. worked out the theory. F. K. designed the samples and performed the experiments. F. K., L.T., A.S. and S.S. analyzed and discussed the results. A.S. and S.S. supervised the project. All authors co-wrote the manuscript.

## Competing Interests Declaration
The authors declare no competing interests.

## Data and Materials Availability Statement
Further/raw data and analyzation code are available on request.

## List of Supplementary Materials
Methods
Supplementary Text
Reference (*23*)

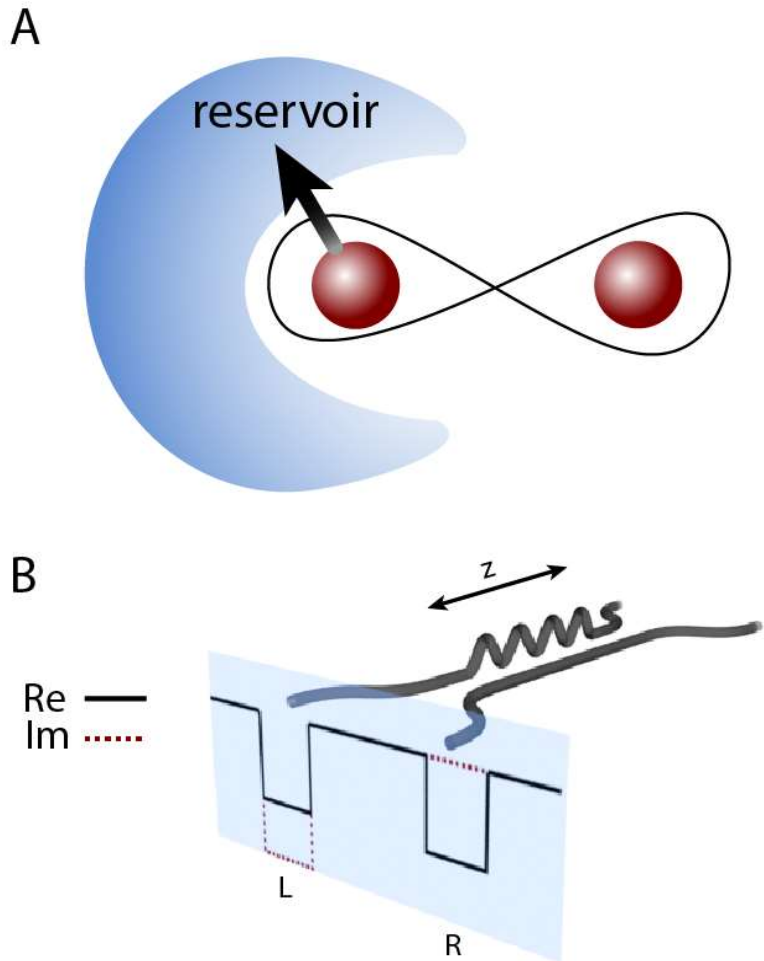

**Fig. 1. Combining photon correlations and PT symmetry.**

(**A**) Realization of a non-Hermitian system: in a two-particle state, one of the partitions is coupled to a reservoir, resulting in Markovian loss. (**B**) Implementation of passive PT symmetry in a coupled two-waveguide system via a symmetric refractive index distribution (real part, black) and an asymmetric loss distribution (imaginary part, red). The left waveguide $L$ is modulated such that it exhibits additional bending losses. The propagation length is $z$.

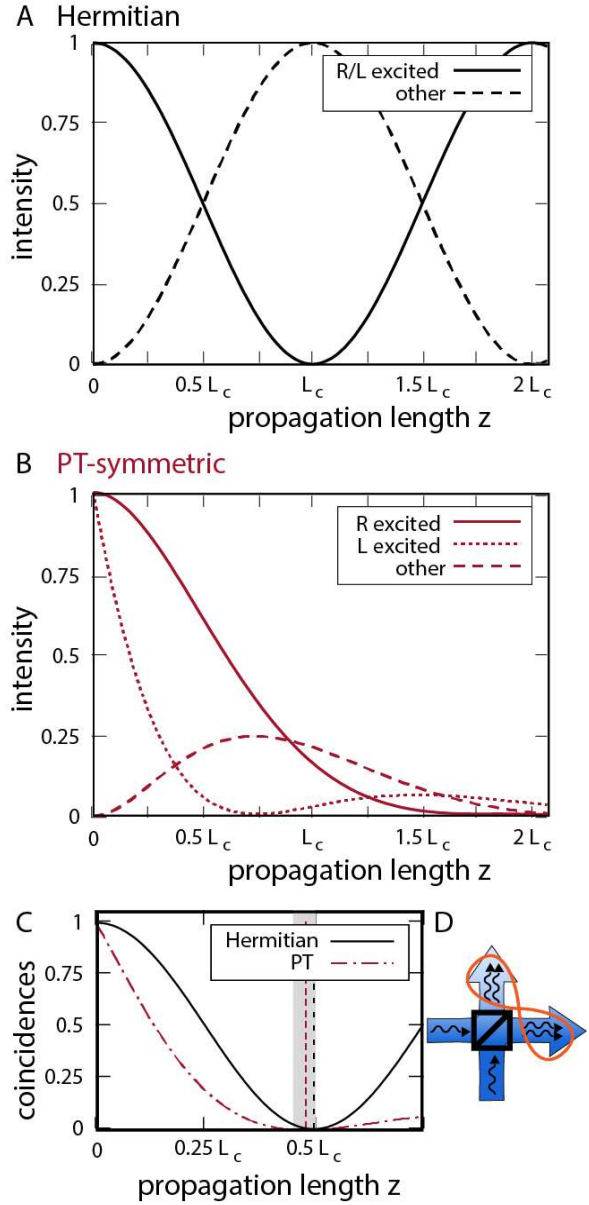

**Fig. 2: Analytical solution of the lossy directional coupler.**

(**A**),(**B**) Intensity vs. propagation length $z$ of the waveguide coupler with a coupling constant $\kappa = 0.25\ cm^{-1}$. (A) In the Hermitian case ($\gamma = 0$) the intensities in both waveguides are equal at half the coupling length $L_c = \pi/(2\kappa)$. (**B**) The evolution in the PT-symmetric case depends on whether light is coupled into the lossy ($L$, dotted line) or lossless waveguide ($R$, solid line). The initially non-excited waveguide intensity evolves independently of the input (dashed line). The loss $\gamma = 0.35\ cm^{-1}$ asymmetrically changes the oscillation. (**C**),(**D**) Two indistinguishable photons are launched into the two waveguides. (**C**) Analytical solution for the coincidence rate $\Gamma$ showing a HOM-Dip. In the Hermitian case (black solid line), the bunching occurs at the 50/50 splitting point $z_H$ (vertical black dashed line). In the lossy PT-symmetric case (red dashed-dotted line), the bunching occurs after a shorter propagation length $z < z_H$, (vertical red dashed line). The PT-symmetry breaking threshold of $\gamma = 2\kappa$ limits the region in which bunching is possible (shaded area).

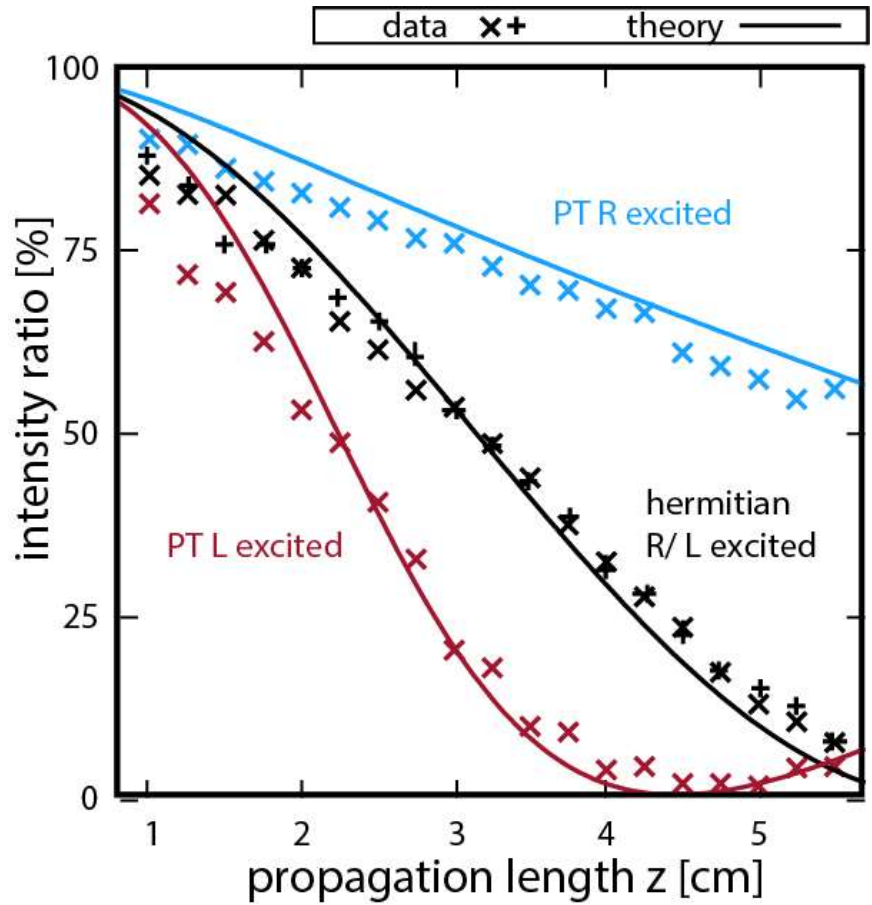

**Fig. 3: Measurement of the intensity ratio for directional couplers at different propagation lengths.**

PT-symmetric couplers and a reference set of corresponding Hermitian couplers are investigated. The intensity ratios are not affected by mean loss, but only by loss asymmetry. Errors are included in marker size, solid lines show theory curves for designed parameters of $\kappa = 0.26\ cm^{-1}$ and $\gamma = 0.2\ cm^{-1}$. For the PT couplers, the light evolution is different for the $R$ (lossless, blue) and $L$ (lossy, red) input. The Hermitian couplers are shown in black.

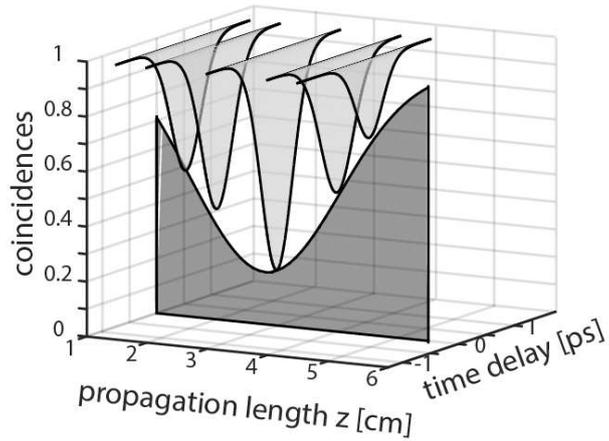
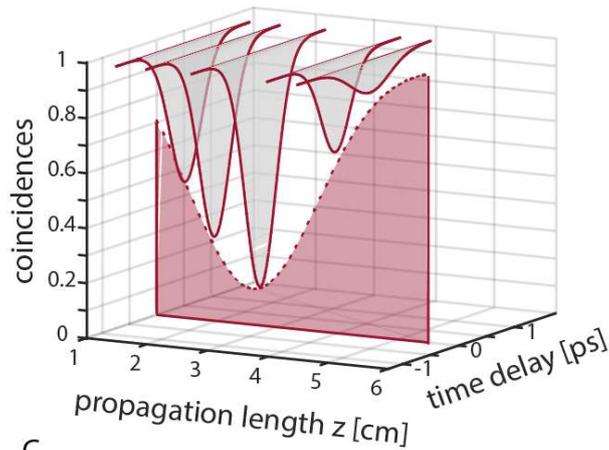
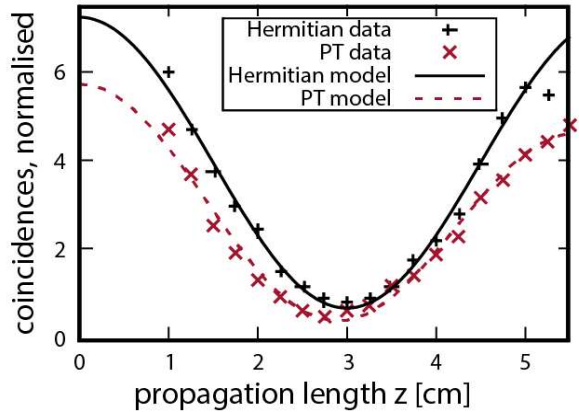

**Fig. 4: Measurement of the HOM dip for a set of PT-symmetric and corresponding Hermitian couplers.**

A pair of photons with a certain time delay is launched into the couplers. (**A**),(**B**) The two-photon coincidence is measured as a function of the time delay, resulting in a HOM dip for each coupler. The depth of the dip depends on the propagation length of the individual coupler. The coincidences have been fitted to Gaussian functions. The HOM dip is found at the minimum coincidence rate for all propagation lengths and time delays. The Hermitian HOM dip shows a visibility of (87 ±2) %, whereas the PT-symmetric HOM dip has a visibility of (90±4) %. (**C**) The HOM dip in the Hermitian case occurs at $z_H = 3.0\ cm$, whereas the HOM dip in the PT-symmetric situation occurs at $z = 2.8\ cm < z_H$.